\documentclass[aps,prb,preprint,showpacs,showkeys,superscriptaddress,amsmath,amssymb,floatfix,author-numerical,nofootinbib]{revtex4-1}

\usepackage{bm}
\usepackage{graphicx}
\usepackage{amsmath}
\usepackage{hyperref}
\usepackage{epsfig}
\usepackage{epstopdf}

\usepackage[ngerman, english]{babel}

\DeclareGraphicsExtensions{.eps}

\usepackage{hyperref}
\hypersetup{%
	colorlinks=true,
	linkcolor=blue,
	urlcolor=blue,
	citecolor=blue
}

\begin{document}

\title{All-nitride and In-free Al$_x$Ga$_{1-x}$N:Mn/GaN distributed Bragg reflectors for the near-infrared}

\author{Giulia Capuzzo}
\email{giulia.capuzzo@jku.at}
\affiliation{Institut f\"ur Halbleiter-und-Festk\"orperphysik, Johannes Kepler University, Altenbergerstr. 69, A-4040 Linz, Austria}

\author{Dmytro Kysylychyn}
\affiliation{Institut f\"ur Halbleiter-und-Festk\"orperphysik, Johannes Kepler University, Altenbergerstr. 69, A-4040 Linz, Austria}

\author{Rajdeep Adhikari}
\affiliation{Institut f\"ur Halbleiter-und-Festk\"orperphysik, Johannes Kepler University, Altenbergerstr. 69, A-4040 Linz, Austria}

\author{Tian Li}
\affiliation{Institute of Physics, Polish Academy of Science, al. Lotnikow 32/46, 02-668 Warsaw, Poland}

\author{Bogdan Faina}
\affiliation{Institut f\"ur Halbleiter-und-Festk\"orperphysik, Johannes Kepler University, Altenbergerstr. 69, A-4040 Linz, Austria}

\author{Aitana Tarazaga Mart\'{\i}n-Luengo}
\affiliation{Institut f\"ur Halbleiter-und-Festk\"orperphysik, Johannes Kepler University, Altenbergerstr. 69, A-4040 Linz, Austria}

\author{Alberta Bonanni}
\email{alberta.bonanni@jku.at}
\affiliation{Institut f\"ur Halbleiter-und-Festk\"orperphysik, Johannes Kepler University, Altenbergerstr. 69, A-4040 Linz, Austria}



  \renewcommand{\abstractname}{}    

  \begin{abstract}

  	Since the technological breakthrough prompted by the inception of light emitting diodes based on III-nitrides, these material systems have emerged as strategic semiconductors not only for the lighting of the future, but also for the new generation of high-power electronic and spintronic devices. While III-nitride optoelectronics in the visible and ultraviolet spectral range is widely established, all-nitride and In-free efficient devices in the near-infrared (NIR) are still wanted. Here, through a comprehensive protocol of design, modeling, epitaxial growth and in-depth characterization, we develop Al$_x$Ga$_{1-x}$N:Mn/GaN NIR distributed Bragg reflectors and we show their efficiency in combination with GaN:(Mn,Mg) layers containing Mn-Mg$_{k}$ complexes optically active in the telecommunication range of wavelengths. 
  	
  \end{abstract}
  
 \maketitle

 \section*{Introduction}
 
  Over two decades ago a series of fundamental breakthroughs in the area of gallium nitride (GaN)-based semiconductor materials has led to the first demonstration of high efficiency and high brightness blue light emitting diodes (LEDs)\cite{Nakamura:1993_APL,Nakamura:1995_JVSTa}. Currently, GaN-based blue and white LEDs reach efficiencies exceeding those of any conventional light source and III-nitride-based heterostructures represent the building-blocks not only of state-of-the-art laser diodes\cite{Yoshida:2008_NP}, blue and white LEDs\cite{Gutt:2012_APE}, but also of high mobility transistors\cite{Mishra:2002_IEEE,Sun:2015_APL}, high power electronic\cite{Shur:1998_SSE,Yeluri:2015_APL} and spintronic\cite{Dietl:2015_RMP} devices. The technological importance of III-nitrides is justified by a number of remarkable properties, including a widely tunable band-gap, the availability of both $n$- and $p$-type material, a remarkable thermal stability and large heat conductivity. In order to extend the functionalities of III-nitride systems to the near-infrared (NIR) range -- $e.g.$ for telecommunication applications -- currently these materials are either doped with rare earths and Er in particular\cite{MacKenzie:1998_APL,Torvik:1997_JAP,Przybylinska:2001_PbCM}, or alloyed with a considerable amount of In\cite{Wu:2009_JAP,Mi:2015_PSSb}, challenging the epitaxial growth and the homogeneity of the layers\cite{Kawakami:2001_JPCM}.
  
  Recently, we have reported that the co-doping of GaN with Mn and Mg results in the formation of robust cation complexes Mn-Mg$_{k}$ [\citenum{Devillers:2012_SR,Devillers:2013_APL}], responsible for a room-temperature (RT) broad IR emission that covers two of the telecommunication windows, respectively centered at 1.33 $\mu$m and 1.55 $\mu$m, opening wide perspectives towards the realization of efficient NIR devices not requiring rare earths or In.
  
  Moreover, by embedding in an optical cavity layers of GaN:(Mn,Mg) containing the Mn-Mg$_{k}$ complexes, a variety of NIR opto-electronic devices, like vertical-cavity surface-emitting lasers (VCSELs)\cite{Carlin:2005_APL,Zhang:2006_APL}, resonant-cavity light emitting diodes (RCLEDs)\cite{Song:2000_APL}, and single photon emitters (SPE)\cite{Yuan:2002_S,Holmes:2014_NL}, can be envisaged.  
  
  Distributed Bragg reflectors (DBR) are essential elements of an optical cavity and while stacks of dielectric materials deposited by electron beam evaporation are well established in the fabrication of DBRs for the NIR range\cite{Convertino:1997_APL,Feng:2013_JVSTa,Tripathi:2015_JMS}, the epitaxial growth of semiconductor-based DBRs by metalorganic vapor phase epitaxy (MOVPE) or by molecular-beam epitaxy (MBE)\cite{Blum:1995_APL,Xia:2005_OLT} is highly desirable, since in this way the optically active layers can be grown directly on top of a buried DBR or sandwiched between two reflectors forming a resonator. Although these epitaxial protocols are widely reported for the ultraviolet (UV)\cite{Kruse:2011_PSSb,Someya:1998_APL,Wang:2004_APL,Diagne:2001_APL} and deep-UV\cite{Mitrofanov:2006_APL,Moe:2006_PSSa,Brummer:2015_APL} range, the development of all-nitride NIR DBR/active region structures is in its infancy.\\ 
  In a DBR the optical stop-band, $i.e.$ the narrow range of wavelengths for which the propagation of light is strongly inhibited, is essentially due to multiple interference processes at the interfaces of a stack consisting of the repetition of two alternating layers -- a Bragg pair -- with respectively low and high refractive index. The separation between subsequent interfaces should be a multilple of a quarter of the design wavelength. The performance of the reflector is determined by (i) the contrast in the refractive index between the two materials of the Bragg pair and (ii) by the number of pairs. 
  Several groups reported on the fabrication of Al$_x$Ga$_{1-x}$N/GaN DBRs in the UV and visible range and in the majority of reports strain engineering comes into play, due to the necessity of overcoming the detrimental effects of the relaxation of stress originating from the lattice mismatch between GaN and its alloys. Among the procedures employed, we recall the use of GaN/Al$_x$Ga$_{1-x}$N or GaN/AlN superlattice (SL) insertion layers to reduce the biaxial tensile strain and to quench the generation of cracks\cite{Huang:2006_APL,Kruse:2011_PSSb,Nakada:2003_JJAP}, useful also in case of thick Al$_x$Ga$_{1-x}$N films\cite{Wang:2002_APL}. Alternative solutions consist in inserting single or multiple AlN interlayers during the growth of the DBR sequence\cite{Waldrip:2001_APL}, supported by an Al$_x$Ga$_{1-x}$N layer or buffer \cite{Natali:2003_APL,Dartsch:2008_JCG,Kruse:2011_PSSb}. 
  We have recently demonstrated that the incorporation of $<$1\% of Mn during the epitaxy of Al$_x$Ga$_{1-x}$N, affects the plastic relaxation of the layers and increases substantially their critical thickness on GaN~[\citenum{Devillers:2015_CGD}]. 
  In this work, we report on the design and fabrication of In-free low-defect Al$_x$Ga$_{1-x}$N:Mn/GaN DBRs grown by MOVPE for the spectral region between 900\,nm and 1500\,nm and on their effect on the NIR emission from a GaN:(Mn,Mg) active layer. 
 

 \section*{Experimental and modeling}  
The samples are grown by MOVPE on 2"\emph{c}-plane sapphire substrates in an AIXTRON 200RF horizontal reactor, according to procedures we have described elsewhere\cite{Stefanowicz:2010_PRB,Devillers:2012_SR,Devillers:2015_CGD}. The precursors employed for Ga, N, Al, Mn and Mg are trimethylgallium (TMGa), ammonia (NH$_3$), trimethylaluminium (TMAl), bis-methylcyclopentadienyl-manganese (MeCp$_2$Mn), and dicyclopentadienyl-magnesium (Cp$_2$Mg) respectively. The deposition process is carried out under H$_2$ atmosphere.  
 After the growth of a Al$_x$Ga$_{1-x}$N nucleation layer at 540\,$^\circ$C and p=200 mbar, the annealing process is carried out at 975\,$^\circ$C. A 1~$\mu$m Al$_{0.12}$Ga$_{0.88}$N:Mn buffer layer is then deposited epitaxially at 975\,$^\circ$C and p=100 mbar.
 Upon deposition of the buffer, the Al$_{0.27}$Ga$_{0.73}$N:Mn/GaN Bragg pairs are grown at the same temperature, and in samples \#B, \#D, \#F and \#H, a 130 nm thick GaN:(Mn,Mg) active layer is deposited at 850\,$^\circ$C and p=200 mbar. For sample \#I the GaN:(Mn,Mg) layer deposition starts directly after the Al$_{0.12}$Ga$_{0.88}$N:Mn buffer, providing a reference without Bragg reflector.

\textit{In situ} and on-line kinetic ellipsometry ensures the direct control of the deposition process and provides information on the thickness of the layers, which is then confirmed by \emph{ex~situ} spectroscopic ellipsometry and transmission electron microscopy (TEM) in both conventional (CTEM) and scanning mode (STEM), performed in a FEI Titan Cube 80-300 operating at 300\,keV and in a JEOL 2010F working at 200\,keV.
 Bright/dark-field (BF/DF), high resolution TEM (HRTEM) and high angle annular dark field (HAADF) are employed for the in-depth structural characterization of the structures, and mapping is performed with energy filtered TEM (EFTEM), at the Al $\textit{K}$ absorption edge.
 Cross-section TEM specimens are prepared by mechanical polishing, dimpling and final ion milling in a Gatan Precision Ion Polishing System.\\
 Information on the morphology of the surface is obtained from atomic force microscopy (AFM) in tapping mode with a VEECO Dimension 3100, while the Al concentration is calculated from the position of the (0002) and $(\overline{1}015)$ diffraction peaks of Al$_x$Ga$_{1-x}$N(:Mn), measured on a PANalytical's X'Pert PRO Materials Research Diffractometer (MRD) equipped with a hybrid monochromator with a 1/4$^{\circ}$ divergence slit. The diffracted beam is measured with a solid-state PixCel detector used as 256-channels detector with a 9.1\,mm anti-scatter slit. 
 Reflectivity measurements are carried out at room temperature with a Bruker VERTEX 80 Fourier-transform IR spectrometer.
 Photoluminescence (PL) spectra are acquired  at 6~K and at room temperature, using a diode laser with an excitation wavelength of 442~nm and an InGaAs line detector. 
 
 The design of the DBRs in this work is supported by reflectivity simulations based on the transfer matrix method (TMM) [\citenum{Mitsas:1995_AO}]. With this formalism, the relation between the electric fields of the incident, reflected and transmitted light is given by modeling the multilayer structure as a series of interfaces and propagation regions represented by a scattering matrix (system transfer matrix), which is the successive product of: (i) the refractive matrices describing the reflection and transmission at a single interface and (ii) the phase matrices accounting for the phase shift caused by the propagation through a layer. Within this model, the whole transmission and reflectance spectrum of an arrangement of dielectric layers can be obtained, once the refractive indices of the involved materials are known.
 For the present work, the refractive indices of Al$_x$Ga$_{1-x}$N:Mn alloys with different Al content and 0.2\% of Mn, have been established by spectroscopic ellipsometry measurements. 
 
 \section*{Results and discussion}
 
 \subsection*{Refractive indices and design}
\begin{figure}[h]
 	\centering
 	\includegraphics[width=0.95\textwidth]{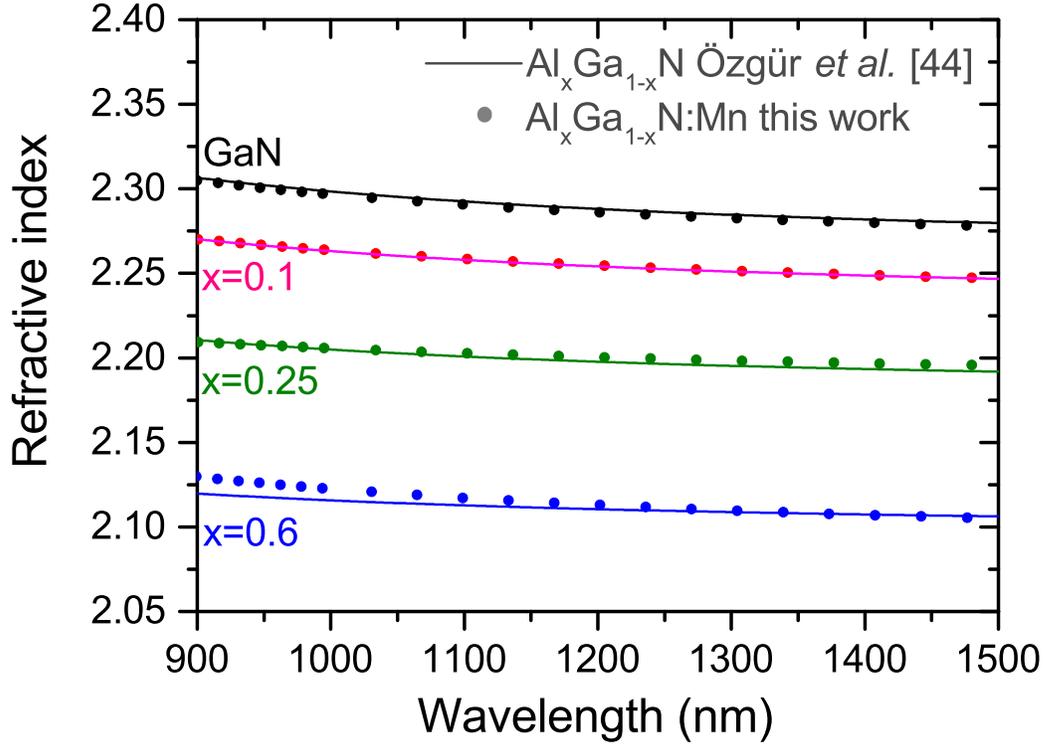}
 	\caption{Refractive indices of Al$_x$Ga$_{1-x}$N:Mn with various Al contents and 0.2\% Mn (dots) compared with those reported by \"Ozg\"ur $et$ $al.$ [\citenum{Ozgur:2001_APL}] for Al$_x$Ga$_{1-x}$N (solid lines), extrapolated for NIR wavelengths.}
 	\label{fig:fig1}
 \end{figure}

 The refractive indices of Al$_x$Ga$_{1-x}$N:Mn alloys with different Al content and 0.2\% of Mn are reported in Fig.~\ref{fig:fig1} together with those extrapolated from the model based on the first-order Sellmeier dispersion formula employed by \"Ozg\"ur $et$ $al.$ [\citenum{Ozgur:2001_APL}] for Al$_x$Ga$_{1-x}$N. The values are shown for wavelengths between 900~nm and 1500~nm, corresponding to the range of NIR emission from the Mn-Mg$_{ k}$ complexes in the GaN:(Mn,Mg) active layer of specific interest\cite{Devillers:2012_SR}.
  
 In order to obtain high reflectivity and a wide stop-band, a significant difference in the refractive indices of the two materials of the Bragg pairs is required, implying that the higher the concentration of Al in the Al$_x$Ga$_{1-x}$N layers, the more pronounced is the optical contrast with the GaN counterpart. However, one must take into consideration, that the critical thickness of Al$_x$Ga$_{1-x}$N on GaN [\citenum{Lee:2004_APL}] decreases dramatically with increasing the Al content $x$ and the strain due to the lattice mismatch is released through the formation of dislocations and eventually through cracking of the structure. On the other hand, as already mentioned, the introduction of as less as 0.2\% of Mn into Al$_x$Ga$_{1-x}$N allows us to increase significantly its critical thickness on GaN. Based on these limitations related to the epitaxial growth of mismatched materials -- but taking advantage of the surfactant effect of Mn -- and having GaN as high refractive index material and Al$_x$Ga$_{1-x}$N:Mn as low refractive index layer, we compromise on a target Al content $x$=0.27. Moreover, we give for a stop-band in the wanted range -- which includes, at 1200~nm, the most intense emission from the Mn-Mg$_{k}$ in the GaN:(Mn,Mg) active layer -- a thickness of 137~nm for the Al$_{0.27}$Ga$_{0.73}$N:Mn layer and of 131~nm for the GaN one. With these values, we show that one can reach already a 60\% of reflectance with a multilayer structure consisting of 20 strained Bragg pairs.
 
\begin{figure}[h]
   	\centering
   	\includegraphics[width=0.85\textwidth]{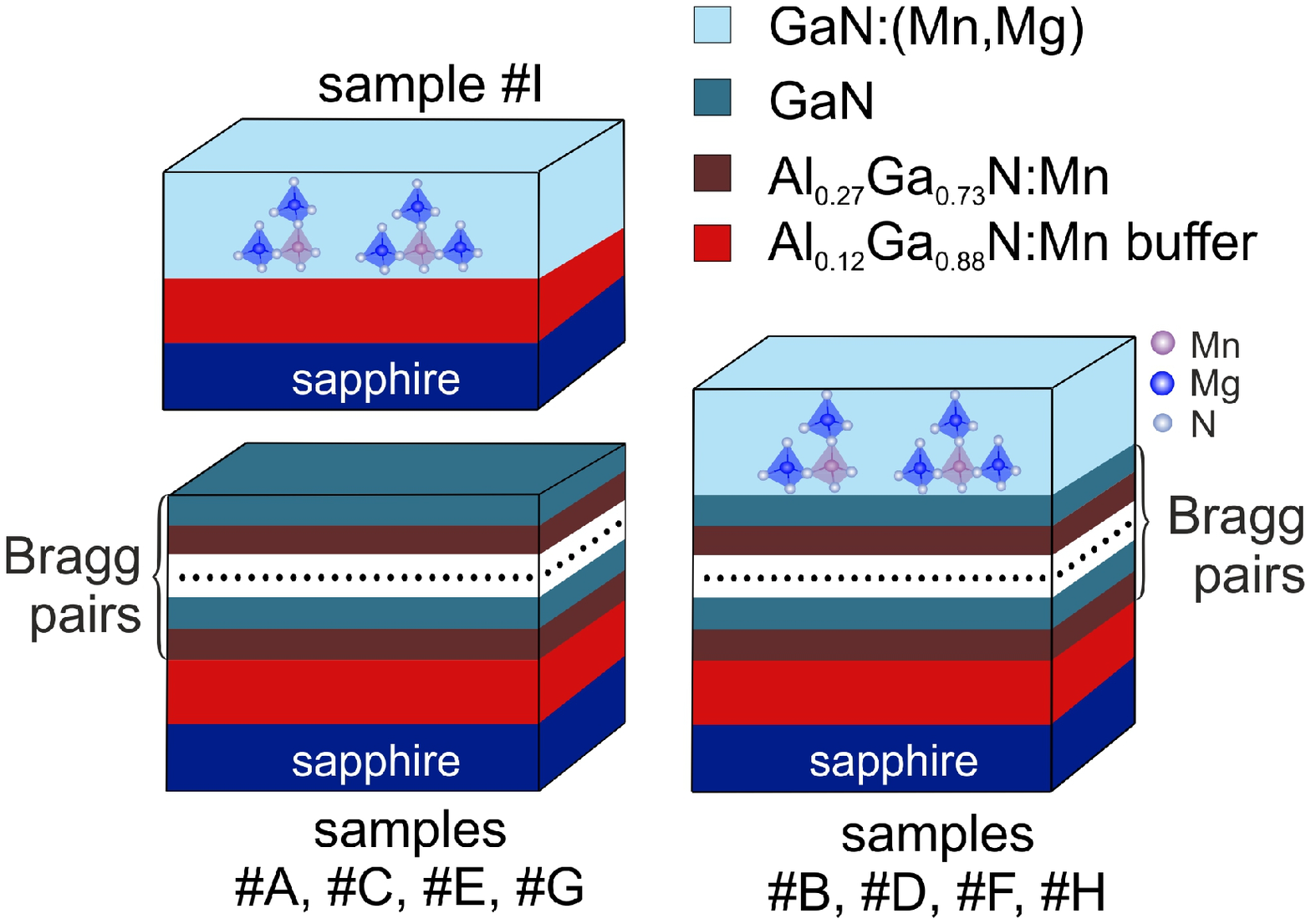}
   	\caption{Sketches of the samples structures. Sample \#I is a reference without DBR and consisting of a GaN:(Mn,Mg) active layer deposited directly on the Al$_{0.12}$Ga$_{0.88}$N:Mn buffer.}
   	\label{fig:fig2}
   \end{figure}

 The schematic model of the studied structures is reported in Fig.~\ref{fig:fig2}, while the number of Bragg pairs for each investigated sample is provided in Table~\ref{table:sample}, together with details on the presence of the active layer.
  
\begin{table}[h]
 	\caption{Investigated samples: number $n$ of Bragg pairs and presence of GaN:(Mn,Mg) active layer. In samples \#B, \#D and \#F, the 1200~nm emission from the Mn-Mg$_{k}$ complexes in the GaN:(Mn,Mg) active layer is in the range of maximum reflectivity of the DBR (stop-band). }
 	\label{table:sample}
 	\begin{tabular}{|c|c|c|}
 		\hline
 		Sample & \begin{tabular}[c]{@{}c@{}}Number $ n $ of Bragg pairs \\ 137 nm/131 nm\\ (Al$_{0.27}$Ga$_{0.73}$N:Mn/GaN)\end{tabular} & \begin{tabular}[c]{@{}c@{}}GaN:(Mn,Mg) \\ active layer \\ (130 nm)\end{tabular} \\
 		
 		\hline
 		\#A 				& 5 			& no			\\
 		\#B					& 5 			& GaN:(Mn,Mg)													\\
 		\#C					& 10 				     								& no																	\\
 		\#D					& 10														& GaN:(Mn,Mg)													\\
 		\#E 				& 20 				     								& no																	\\
 		\#F					& 20 														& GaN:(Mn,Mg)													\\
 		\#G					& 20 				(112 nm/112 nm) 		& no				  												\\
 		\#H					& 20 				(112 nm/112 nm) 		& GaN:(Mn,Mg)													\\
 		\#I					& 0 				& GaN:(Mn,Mg)																							\\
 		\hline
 	\end{tabular}
\end{table}

 \subsection*{Towards an optimized DBR}
\begin{figure} [h]
  	\centering
  	\includegraphics[width=0.75\textwidth]{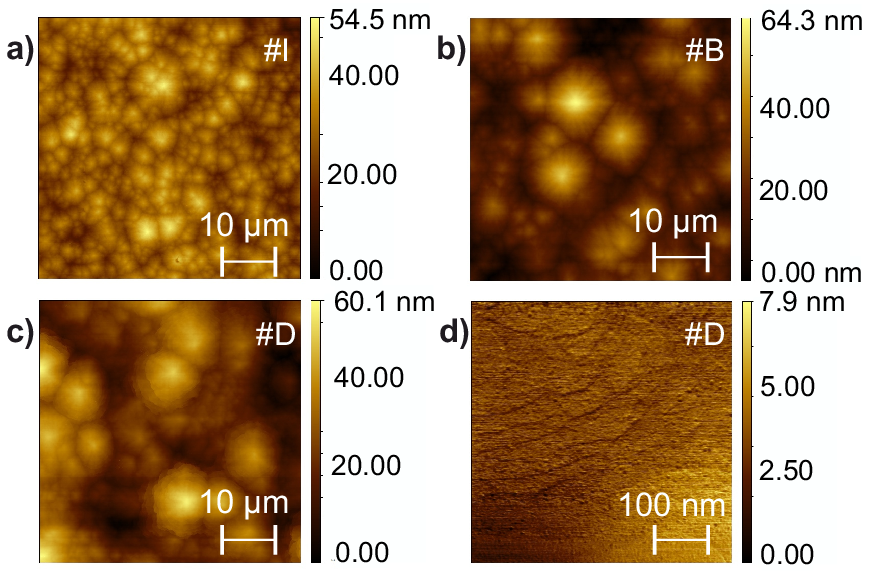}
  	\caption{Atomic force micrographs for: a) sample \#I (reference without DBR) , b) sample \#B (5-fold DBR and GaN:(Mn,Mg) active layer), c) and d) sample \#D (10-fold DBR and GaN:(Mn,Mg) active layer). On the larger scale a)-c), the features typical of Mn-doped samples are visible, while in the 1~$\mu$m-scale picture d) atomic terrace edges characteristic of a step-flow growth mode can be distinguished.}
  	\label{fig:fig3}
  \end{figure}

  A protocol of in-depth post-growth characterization of the structures is employed in order to establish the relation between growth parameters, crystallographic arrangement, chemical composition and optical response of the investigated structures. 
  On the large scale, the surface of all the samples studied by atomic force microscopy (AFM) and reported in Fig.~\ref{fig:fig3} a), b) and c) shows a morphology already observed in the Al$_x$Ga$_{1-x}$N:Mn samples studied by our group recently \cite{Devillers:2015_CGD}. In the presence of the GaN:(Mn,Mg) active layer and with increasing number of Bragg pairs, the average size (both in-plane and in the growth direction) of the surface features increases, as seen when comparing the reference sample \#I (active layer directly deposited on the buffer) in Fig.~\ref{fig:fig3} a) with Fig.~\ref{fig:fig3} b) and c), where a 5-fold and a 10-fold DBR have been added, respectively. In the high resolution images, on the other hand, it is possible to distinguish the atomic terrace edges characteristic of a step-flow growth mode, as evidenced in Fig.~\ref{fig:fig3} d). 
 
\begin{figure}[h]
   	\centering
   	\includegraphics[width=0.5\textwidth]{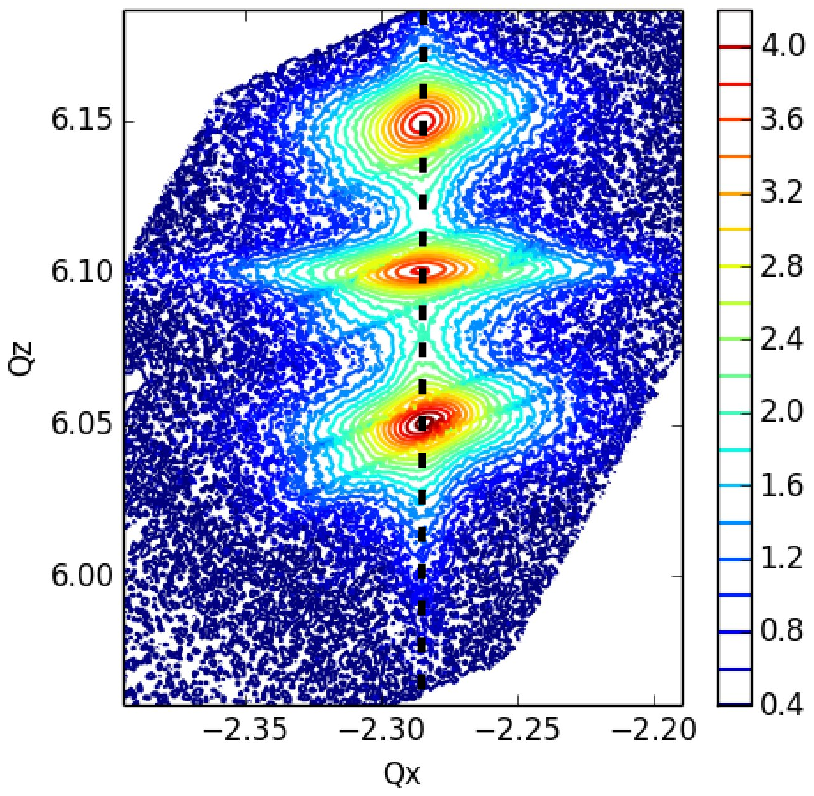}
   	\caption{Reciprocal space maps around GaN and Al$_x$Ga$_{1-x}$N(:Mn) ($\bar{1}$015) for sample \#D. The structure is grown pseudomorphically on the Al$_{0.12}$Ga$_{0.88}$N:Mn buffer. The intensity is reported in logarithmic scale. A vertical dashed line along the GaN $(\overline{1}01l)$ is drawn as guide to the eye.}
   	\label{fig:fig4}
   \end{figure}

   X-ray diffraction reciprocal space maps (RSMs) measured about the ($\bar{1}$015) reflection of GaN and Al$_x$Ga$_{1-x}$N(:Mn) show that all the structures under investigation are grown pseudomorphically. The RSM of sample \#D is reported in Fig.~\ref{fig:fig4}, where low, intermediate and high Q$_z$ values correspond to the GaN layer, Al$_{0.12}$Ga$_{0.88}$N:Mn buffer and Al$_{0.27}$Ga$_{0.73}$N:Mn layers respectively. The centers of the three peaks are aligned at the same value of Q$_x$, pointing to a strained state of the films.  
   The Al content in the layers is quantified from the position of the $(\overline{1}015)$ peak and according to the Vegard's law satisfied by the considered compounds\cite{Devillers:2015_CGD}. The obtained concentrations are confirmed through energy-dispersive x-ray spectroscopy (EDX) measurements, are similar for all the samples in the series, and correspond to $(12.0\pm1.0)$\% in the buffer and $(26.9\pm1.0)$\% in the Al$_x$Ga$_{1-x}$N:Mn Bragg layers, respectively. The Mn content is $<$0.2\% (cations) in the buffer and in the Bragg layers.

\begin{figure}[h]
 	\centering
 	\includegraphics[width=0.65\textwidth]{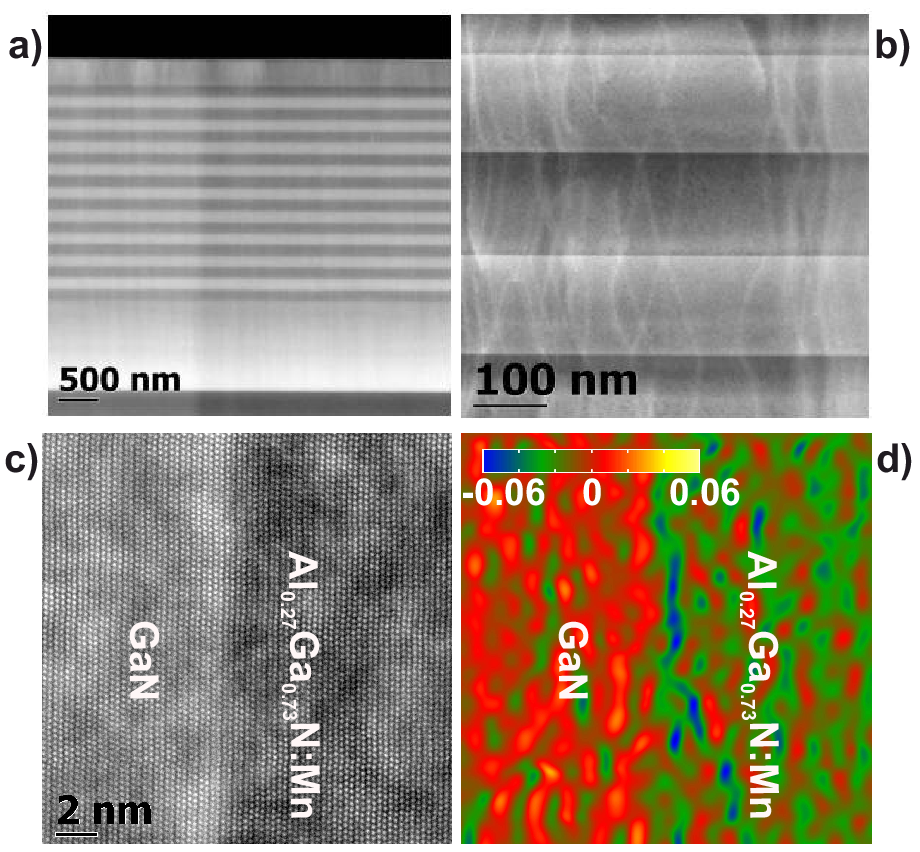}
 	\caption{Panels a) and b): HAADF/STEM of sample \#D. Light and dark alternating areas correspond to the GaN and Al$_x$Ga$_{1-x}$N:Mn regions of the Bragg pairs, respectively. In panel a) the 130~nm thick GaN:(Mn,Mg) active layer is also distinguishable at the top of the structure. Panel c) high resolution HAADF/STEM of sample \#H acquired along the [11$\overline{2}$0] zone axis, with atomically defined interface between the GaN and Al$_x$Ga$_{1-x}$N:Mn layers of one Bragg pair. Panel d) GPA strain mapping for the interface reported in panel c).}
 	\label{fig:fig5}
 \end{figure}

 The transmission electron microscopy (TEM) analysis of the structures points to the absence of major defects such as cracks or V-shaped ones in the heterostructures. Light and dark alternate regions in the scanning TEM (STEM) image reported in Fig.~\ref{fig:fig5} a) for sample \#D correspond to GaN and Al$_x$Ga$_{1-x}$N:Mn Bragg layers, respectively, while the defined Z-contrast in the high resolution high angle annular dark field (HAADF)/STEM image of Fig.~\ref{fig:fig5} c) is an indication of the atomically sharp interface between the Al$_x$Ga$_{1-x}$N:Mn and the GaN layers. The thickness of the single layers is in accord with the nominal one expected from the growth parameters and with those required by the TMM model. 
 The strain state around the Al$_x$Ga$_{1-x}$N:Mn/GaN interface already evidenced by x-ray diffraction (XRD) is confirmed by the geometric phase analysis (GPA) reported in Fig.~\ref{fig:fig5} d) for an interface in sample \#H. In the studied area, the average strain of the \textit{c}-parameter for the Al$_x$Ga$_{1-x}$N:Mn layer with respect to the GaN layer is -0.012, while its \textit{a}-parameter is matched to the one of the GaN layer, resulting in a pseudomorphic growth of the Bragg pairs and of the whole DBR structure.\\

  \begin{figure}[ht]
  	\centering
  	\includegraphics[width=1\textwidth]{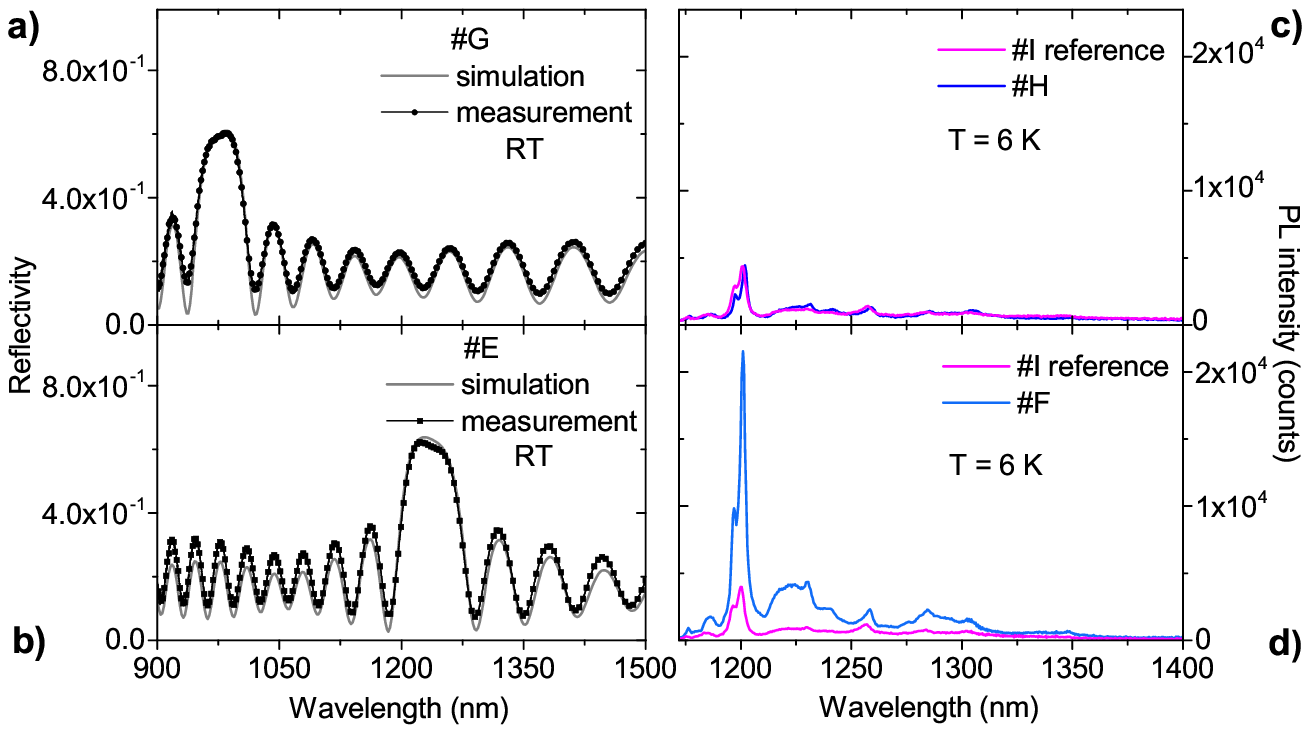}
  	\caption{Left panels: room temperature - measured and calculated - reflectivity of samples \#G – a) and \#E – b) with 20 Bragg pairs confirming the tunability of the stop-band with respect to the Bragg layers thickness. Right panels: low temperature PL comparing sample \#H - c) and \#F - d) with the reference sample \#I. A significant enhancement of the PL signal when the stop-band is closer to 1200\,nm is observed.}
  	\label{fig:fig6}
  \end{figure}

 \subsection*{Effect of the Al$_x$Ga$_{1-x}$N:Mn/GaN  DBR on the optical response of the active layer}
 The simulations performed by applying the TMM method, prove to be a powerful tool to predict the effect of the Bragg layer and Bragg pair thickness on the position of the stop-band of the DBR. By changing the thicknesses of the Al$_{0.27}$Ga$_{0.73}$N:Mn and GaN films, the position of the stop-band in the reflectivity spectra can be finely tuned and $e.g.$ a decrease in the thickness (of the single Bragg layers and, consequently, of the Bragg pair) shifts the center of the stop-band towards low wavelengths. For example -- as evidenced in Fig.~\ref{fig:fig6} a) -- for sample \#G, which consists of 20 Bragg pairs, each of them having 112~nm-thick Al$_{0.27}$Ga$_{0.73}$N:Mn and GaN layers, the center of the stop-band is at 980~nm. On the other hand, an increment of the Al$_{0.27}$Ga$_{0.73}$N:Mn and GaN layers thickness - as in sample \#E - shifts the center of the stop-band to 1235~nm, as seen in Fig.~\ref{fig:fig6} b). The measured reflectivity spectra are in agreement with the simulations carried out for the multilayer structure with the thicknesses discussed above.
 The effect of shifting the stop-band affects the photoluminescence (PL) signal around 1200~nm for the samples having a GaN:(Mn,Mg) active layer on top of a DBR structure. For example, the DBR in samples \#G and \#H generates a stop-band centered at 980~nm, as previously discussed, and consequently, no enhancement in the around-1200~nm PL intensity (no effect of the Bragg reflector) is observed for sample \#H (with GaN:(Mn,Mg) active layer) compared to sample \#I (reference GaN:(Mn,Mg) active layer without DBR), as evidenced in Fig.~\ref{fig:fig6} c). In contrast, sample \#F (with GaN:(Mn,Mg) active layer), having a DBR with stop-band centered around 1200~nm, shows a Mn-Mg$_{ k}$-related PL intensity at 1200\,nm which is at least five times greater that the one from sample \#I (reference GaN:(Mn,Mg) active layer without DBR), as highlighted in Fig.~\ref{fig:fig6} d).
 \newline
The measurements of reflectivity are carried out at room temperature. By considering the changes in the band-gap as a function of temperature for Al$_x$Ga$_{1-x}$N(:Mn) and GaN [\citenum{Morkoc:2009_bookV1}], at 6~K we expect a $\sim$5~nm shift of the DBR stop-band center for the 137~nm/131~nm Al$_x$Ga$_{1-x}$N(:Mn)/GaN Bragg pairs, with a consequent increase of the reflectance at 1200~nm: a similar effect has been reported from PL and reflectivity on Al$_x$Ga$_{1-x}$N/Al$_y$Ga$_{1-y}$N DBRs for the UV range\cite{Wang:2004_APL}.
 
 \begin{figure}[h]
 	\centering
 	\includegraphics[width=0.8\textwidth]{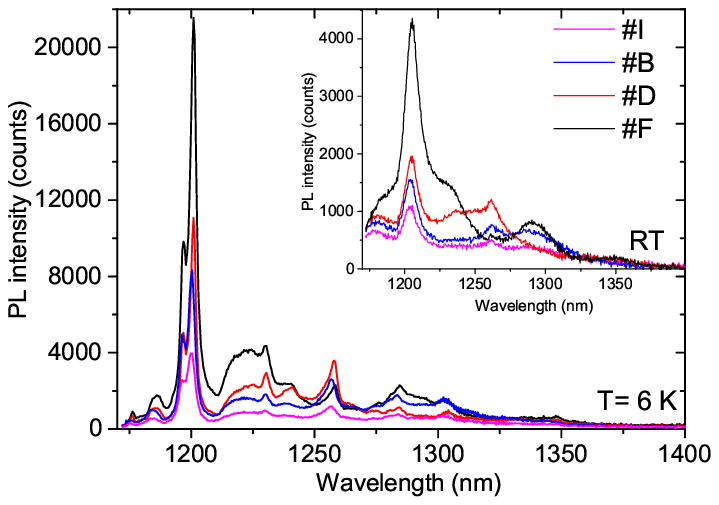}
 	\caption{PL measurements for samples \#B (5 Bragg pairs and active layer), \#D (10 Bragg pairs and active layer), \#F (20 Bragg pairs and active layer) and for the reference \#I (active layer without Bragg pairs) at T=6~K and at room temperature (inset).}
 	\label{fig:fig7}
 \end{figure}
 The full set of PL measurements at 6~K for samples \#B, \#D, \#F and for the reference \#I is reported in Fig.~\ref{fig:fig7}.
 As expected, the intensity of the main Mn-Mg$_{ k}$-related peak at 1200~nm increases as a function of the number of pairs in the DBR. By comparing the PL intensity of reference sample \#I with the one of sample \#F, where the 20-fold DBR has been added, there is an increment of about 5.4 times in the measured intensity. This effect persists up to room temperature, where the intensity of the PL spectra is lower but preserves a systematic dependence on the number of Bragg pairs in the DBR, as reported in the inset to Fig.~\ref{fig:fig7}.

\section*{Synopsis}

All-nitride In-free Al$_x$Ga$_{1-x}$N:Mn/GaN based DBR structures for the NIR range have been designed, fabricated and tested in combination with GaN:(Mn,Mg) layers optically active in the telecommunication range of wavelengths. Simulations based on the TMM method provide an indispensable tool to design and tune the thickness of the various layers constituting the investigated heterostructures.
Photoluminescence measurements up to room temperature reveal the enhancement of the emission intensity from Mn-Mg$_k$ complexes in a GaN:(Mn,Mg) layer grown on the DBR structure, opening up concrete perspectives for the realization of a NIR nitride-based In-free laser.
As the technology for quantum light sources evolves, the development of single photon emitters becomes an essential stage on the roadmap of nitride-based devices\cite{Zhu:2016_EPL,Jarjour:2009_PSSa}. The zero-dimensional nature of the Mn-Mg$_{ k}$ cation complexes -- which identifies them as solotronic objects -- in GaN:(Mn,Mg), together with their structural stability and in combination with tunable Al$_x$Ga$_{1-x}$N:Mn/GaN DBRs paves the way for the design and fabrication of nitride-based single-photon sources.\cite{Shubina:2015_SR}.


\section*{Acknowledgements}

This work was supported by the European Research Council (ERC grant 227690 - FunDMS), by the EC's Horizon2020 Research and
Innovation Programme (grant 645776), by the Austrian Science Foundation --
FWF (P22477 and P24471 and P26830), and by the NATO Science for Peace Programme (Project No. 984735)

\bibliographystyle{naturemag}


\end{document}